\def\prd{Phys. Rev. D}

\def\apj{Astrophys. J.}
\def\apjl{Astrophys. J. Lett.}

\def\mnras{Mon. Not. R. Astr. Soc.}
\def\aap{Astr. Astrophys.}

\def\jcap{JCAP}

\def\qjras{Quarterly Journal of the Royal Astronomical Society}
\def\araa{Annual Review of Astronomy \& Astrophysics}
\newcommand{\ra}{\;\raise1.0pt\hbox{$'$}\hskip-6pt\partial\;}
\newcommand{\lo}{\;\overline{\raise1.0pt\hbox{$'$}\hskip-6pt\partial}\;}

\def \<{\langle}
\def \>{\rangle}

\RequirePackage{snapshot}

\def\useJCAP{aaa}

\ifx\useJCAP\defined
\documentclass[a4paper,11pt]{revtex4}
\usepackage{jcappub}
\fi

\ifx\usePRL\defined
\documentclass[twocolumn,amsmath,amssymb,floatfix,superscriptaddress,showkeys]{revtex4}
\fi

\usepackage{graphicx,epsfig,natbib,color,times,bm,amsmath,multirow,}
\usepackage[ddmmyyyy,hhmmss]{datetime}

\begin{document}
\title{Towards understanding the Planck thermal dust models}

\ifx\useJCAP\defined
\author[a,b]{ Hao Liu} 
\author[a]{Sebastian von Hausegger}
\author[a]{Pavel Naselsky}
\affiliation[a]{Niels Bohr Institute \& Discovery Center, Blegdamsvej 17, DK-2100 Copenhagen, Denmark}
\affiliation[b]{Key laboratory of Particle and Astrophysics, Institute of High Energy Physics, CAS, 19B YuQuan Road, Beijing, China}
\emailAdd{liuhao@nbi.dk}
\fi

\ifx\usePRL\defined
\author{ Hao Liu} \email[mail to: ]{liuhao@nbi.dk}
\affiliation{Niels Bohr Institute \& Discovery Center, Blegdamsvej 17, DK-2100 Copenhagen, Denmark}
\affiliation{Key laboratory of Particle and Astrophysics, Institute of High Energy Physics, CAS, 19B YuQuan Road, Beijing, China}
\author{Sebastian von Hausegger}
\affiliation{Niels Bohr Institute \& Discovery Center, Blegdamsvej 17, DK-2100 Copenhagen, Denmark}
\author{Pavel Naselsky}
\affiliation{Niels Bohr Institute \& Discovery Center, Blegdamsvej 17, DK-2100 Copenhagen, Denmark}
\fi

\ifx\usePRL\defined
\begin{abstract}
\fi

\ifx\useJCAP\defined
\abstract{
\fi

Understanding the properties of dust emission in the microwave domain is an important premise for the next generation of cosmic microwave background (CMB) experiments, devoted to the measurement of the primordial $B$-modes of polarization. In this paper, we compare three solutions to thermal dust emission by the Planck Collaboration~\cite{PlanckDust03,planck_16,planck_com} to point out significant differences between their respective parameters (the spectral index $\beta$, the optical depth $\tau$ and the dust temperature $T_d$). These differences originate from e.g. the priors on the parameters or the contribution of the Cosmic infrared background (CIB). In addition to investigating the angular distributions and statistical properties of each of the $\beta$, $\tau$ and $T_d$-maps for the whole sky, we also compute cross-correlations among the maps, specifically the $\beta-T_d$ and $\tau-T_d$ correlations. All power spectra differ noticeably from each other, which we claim is partly due to the influence of the CIB. Peculiar behavior in the cross-correlations at dust temperatures $\gtrsim21\,K$ supports this claim; the precise differences depend on the particular solutions considered.
Finally, by the example of two zones on the sky (the BICEP2 zone and a region around the North Celestial Pole), we show that not only the properties of dust are different in these regions on the sky, but moreover the dust emission products do not agree. Furthermore, it is illustrated that the use of average values for dust parameters in one zone will not necessarily be applicable to another zone. In this context, we therefore recommend pixel-based approaches for future analyses, with less stringent constraints in form of priors, despite its higher computational expenditure, and an inclusion of a CIB treatment, which finally allows for a direction dependent removal of dust foregrounds.
The central statement of this brief analysis is that while all available solutions are in rough agreement at $\sim5-20\%$, further progress must be made to match the goals of planned $B$-mode experiments.

\ifx\useJCAP\defined
}
\fi

\ifx\usePRL\defined
\end{abstract}
\fi

\maketitle

\section{Introduction}
\label{sec:introduction}
The next generation of the Cosmic Microwave Background (CMB) experiments, such as LiteBird~\cite{litebird1,litebird2,litebird3}, the CMB-S4 missions~\cite{s4}, CORE~\citep{CORE01,CORE02}, DeepSpace~\citep{deepspace}, PIXIE~\citep{Kogut11}, and Polarbear~\cite{polarbear1,polarbear2} will attempt even more precise measurements of the CMB than available so far, targeting a detection of primordial $B$-modes. In combination with the ongoing LIGO experiments, these missions seek to complement gravitational wave astronomy with cosmological gravitational waves from the very beginning of the evolution of the cosmic plasma, thereby opening up new possibilities to understand the properties of gravity and matter in extreme conditions.

The current status of the search for cosmological gravitational waves can be summarized by an upper bound on the tensor-to-scalar ratio, $r$, measured to be $r<0.12$ at 95$\%$ confidence at a pivot scale of $0.05\,\text{Mpc}^{-1}$~\cite{planck15param}. To allow for a tensor-to-scalar ratio as low as $r\sim10^{-3}-10^{-4}$, predicted by e.g. Higgs driven inflation~\cite{higgsinf} or Starobinsky inflation~\cite{starinf}, drastic technological advancement in the sensitivity of polarization detectors is required. Furthermore, it must not be forgotten, that simultaneous progress in the separation of CMB and foregrounds is prerequisite for a sensible interpretation of the resulting $B$-modes.

Future $B$-mode CMB experiments are designed to cover mainly the frequency range around 100 GHz\footnote{Different experiments measure or are planned to measure in different frequency ranges, e.g. LiteBird in the range 50-320~GHz, the CMB-S4 missions most likely in the range 35-250~GHz, CORE in 45-795~GHz, DeepSpace currently measures at 10-15~GHz but will be extended to include frequencies up to 143~GHz, PIXIE is planned to measure from 30~GHz up to 6 THz, and Polarbear currently measures at 150~GHz.}, where the relative intensity of the CMB is known to be highest. One of the most dominant foreground components in this range is thermal dust emission. Naturally, to be sensitive to values of $r$ as quoted above, the contribution of synchrotron emission and at presumably lower polarization fractions free-free emission, CO-line emission, and Anomalous Microwave Emission must not be neglected either. Furthermore, one should bear in mind alternatives to the single-component thermal dust model~\citep{Fin99,Meisner15}, as well as extensions to given parameterizations for the inclusion of averaging effects, spatially as well as along the line-of-sight~\citep{Chluba2017}.

In this paper we revisit three state-of-the-art solutions to thermal dust emission, namely Planck's 2013 full sky model of the first public release~\cite{PlanckDust03} (hereafter P13), Planck's Commander solution from the second data release~\cite{planck_com} (hereafter C15), and a CIB-free solution derived in the so called GNILC framework~\citep{GNILC00} also from Planck's second data release~\cite{planck_16} (hereafter P16). Differences of the methods used for obtaining these solutions are briefly described below.

In this note, we restrict ourselves to intensity measurements only, an analysis of thermal dust polarization products must await the advent of the future data. However, observations in intensity already reveal a lot about the model's parameters. We focus on highlighting the differences of the parameters derived in the three different approaches and their relation to each other. After a first comparison of the power spectra of the parameters of P13 and P16 which both are provided at high resolution, we justify a smoothing of the maps in section~\ref{sec:tdm}. In section~\ref{sec:compare} we then proceed to compare all parameter maps, including those of C15, at a common resolution of $2^\circ$, regarding their visual appearance, their power spectra, their distributions and their cross-correlations. We identify those areas on the maps with $|b|\gtrsim50^\circ$ to be the main cause for the discrepancies, precisely the regions in which the CIB begins to dominate. In addition, in section~\ref{sec:local}, we contrast two zones on the sky, namely the BICEP2 zone, which has been on everyone's lips in the past, and a region along the north celestial pole (NCP) representative for many ground-based experiments, such as QUIJOTE~\cite{QUIJOTE1,QUIJOTE2,QUIJOTE3,QUIJOTE4,QUIJOTE5,QUIJOTE6} or DeepSpace\cite{deepspace}. We summarize in section~\ref{sec:sum}.

\section{Smoothing of the Dust Model parameter maps}
\label{sec:tdm}

In the optically thin limit, the spectral energy density of thermal dust emission at frequency $\nu$ and in direction ${\bf n}$ is modeled as a so called modified black-body (MBB)~\cite{Camb01,Draine03,DraineLi07,Hildebrand83,PlanckDust03}:
\begin{align}
I_\nu({\bf n})=\tau_0({\bf n})\left(\frac{\nu}{\nu_0}\right)^{\beta({\bf n})} B_\nu(T_d({\bf n}))
\label{eq1}
\end{align}
where $\tau_0({\bf n})$ is the dust optical depth at a reference frequency $\nu_0$, and $\beta({\bf n})$ is the dust spectral index.
\begin{align}
B_\nu(T_d({\bf n}))=\frac{2h\nu^3}{c^2}\left(e^{\frac{h\nu}{kT_d({\bf n})}}-1\right)^{-1}
\label{eq:planck}
\end{align}
is the Planck black-body function with dust temperature $T_d({\bf n})$.

The $\beta({\bf n})$, $T_d({\bf n})$ and $\tau_0({\bf n})$ maps of the three solutions to be compared here have different angular resolution and angular smoothing. In specific, the C15 maps have been smoothed by a $1^\circ$ Gaussian beam, significantly larger than the $5'$ those of P13 and P16 have. In order to make a comparison between the three solutions possible, their angular smoothing must agree.

\begin{figure}[!htb]
\includegraphics[width=0.5\textwidth]{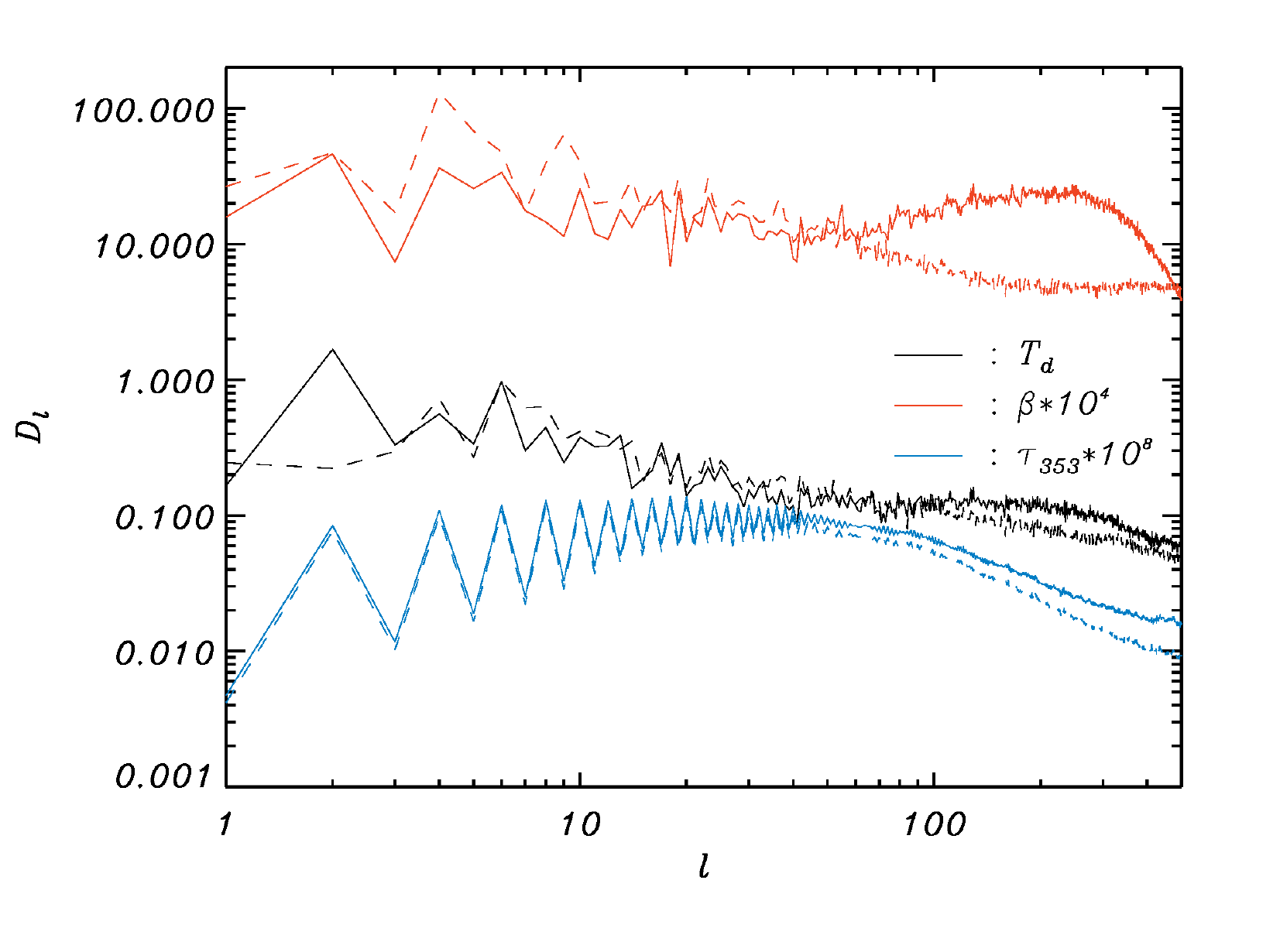}
\caption{The power spectra, $D^X(l)$, as defined in eq.~\ref{eq4}, for $\tau$ (blue), $\beta$ (red), and $T_d$ (black) of P13 (solid lines) and P16 (dashed lines). Note, that for sake of visualization the power spectra $D^\beta(\ell)$ and $D^\tau(\ell)$ have been amplified by factors of $10^4$ and $10^8$, respectively.}
\label{fig1}
\end{figure}

Due to the strong non-linearity of the thermal dust model, eq.~\ref{eq1}, this transition must be done with care. The most straightforward would be to degrade the angular resolution of all intensity maps used in the respective thermal dust fits down to $\Theta=2^\circ$ by applying a Gaussian filter $G\left[(\textbf{n}-\textbf{n'})^2/2\Theta^2\right]$ and subsequently repeat the parameter fits. Unfortunately, this procedure is extremely complex and should not be done here. Instead, we argue, that even resulting, higher resolution maps of $\tau(\textbf{n}),\beta(\textbf{n})$ and $T_d(\textbf{n})$ can be smoothed in regions where their variation is small compared to the mean value or monopole. For this, consider the following decomposition of the respective parameter maps into spherical harmonics $Y_{lm}(\textbf{n})$:
\begin{align}
\left\{
\begin{array}{ccc}
\tau(\textbf{n})\\ \beta(\textbf{n})\\ T_d(\textbf{n})
\end{array}
\right\} = \sum_{l=0}^{\infty}\sum_{m=-l}^{l}
\left\{
\begin{array}{ccc}
\tau_{lm}\\ \beta_{lm}\\ T_{d,lm}
\end{array}
\right\}Y_{lm}(\textbf{n})
\label{eq3}
\end{align}
where $\tau_{lm}, \beta_{lm},T_{d,lm}$ are the corresponding harmonic coefficients.

We compute the power spectrum $D(\ell)$ for each of the parameters as:
\begin{align}
D^X(l)=\frac{l(l+1)}{2\pi(2l+1)}\sum_{-m}^m|X_{lm}|^2
\label{eq4}
\end{align}
where $X$ stands for $\tau, \beta$ or $T_{d}$. The monopoles for the respective parameters are computed to be \mbox{$\bar\tau^{P13}\simeq 1.97 \cdot 10^{-5}$}, \mbox{$\bar\beta^{P13}\simeq1.61$} and $\bar{T}^{P13}_{d}\simeq 19.7\,K$ for P13, and \mbox{$\bar\tau^{P16}\simeq 1.89\cdot10^{-5}$,} \mbox{$\bar\beta^{P16}\simeq1.60$} and $\bar T^{P16}_{d}\simeq 19.4\,K$ for P16. The standard deviations of $\beta$ and $T_d$ lie around $6\%$ of their mean values for both P13 and P16. In Fig.~\ref{fig1} we show the full sky power spectra for all three parameters, from both P13 and P16. Note, that the power spectra $D^\beta(\ell)$ and $D^\tau(\ell)$ were amplified for the sole purpose of visualization. Already here we want to highlight that, while the respective monopoles hardly differ, the multipole dependence of $D^X(\ell)$ reveals a significant mismatch, mainly concentrated at multipoles $\ell\gtrsim50$, especially distinct for the spectral index. These differences arise from the increasing contribution of the CIB at the North and South polar caps, $|b|\gtrsim50^\circ$, as we shall see in the next section. For now, we point to the fact that the amplitudes of the power spectra $D^X(\ell)$ of all maps are smaller than the corresponding monopoles\footnote{$\tau$ is an exception here because its variation can be orders of magnitude higher than the average value; $\tau$ is linear in the dust model, thus the separation in mean value and variation always works regardless of its amplitude.}. For this reason, the intensity of dust emission eq.~\ref{eq1} can be represented as a first order approximation,
\begin{align}
&I_\nu(\textbf{n})\simeq \overline\tau\left(\frac{\nu}{\nu_o}\right)^{\overline\beta}B_{\nu}\left(\overline T_d\right)\times\label{eq5}\\
&\times\left\{1+\frac{\Delta\tau(\textbf{n})}{\overline\tau}+\Delta\beta(\textbf{n})\cdot\ln\left(\frac{\nu}{\nu_0}\right)+\left.\frac{d\ln B_{\nu}}{d\ln T_d}\right|_{\overline T_d}\cdot\frac{\Delta T_d(\textbf{n})}{\overline T_d}\right\}.\nonumber
\end{align}

As Gaussian smoothing is applied to a map via a linear operator, we now see from eq.~\ref{eq5}, that it is equivalent to smooth the intensity map as suggested before and to smooth the individual parameter maps $\tau(\textbf{n}),\beta(\textbf{n})$ and $T_d(\textbf{n})$ afterwards. Note that the problem of extracting physically meaningful parameters in averaged regions on the sky (via instrumental beams, or data processing) has also been discussed in more details in~\citep{Chluba2017}. While we present above considerations mainly to justify the smoothing of the parameter maps, the authors of \citep{Chluba2017} go further and not only compute general expansions (beyond first order) of spectral emission distributions but also present solution approaches to beam and line-of-sight averaging effects.

\section{Full-sky comparison of the solutions}
\label{sec:compare}

\begin{figure*}[!htb]
\centering
\hbox{
\centerline{\includegraphics[width=0.32\textwidth]{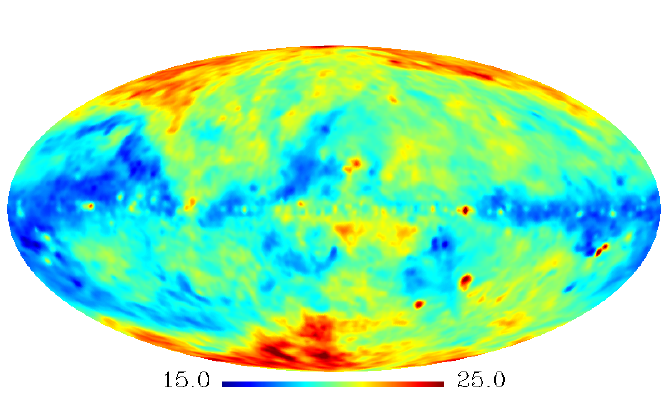}
\includegraphics[width=0.32\textwidth]{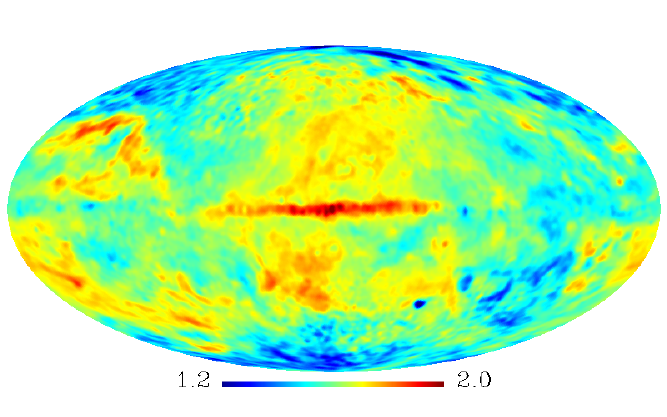}
\includegraphics[width=0.32\textwidth]{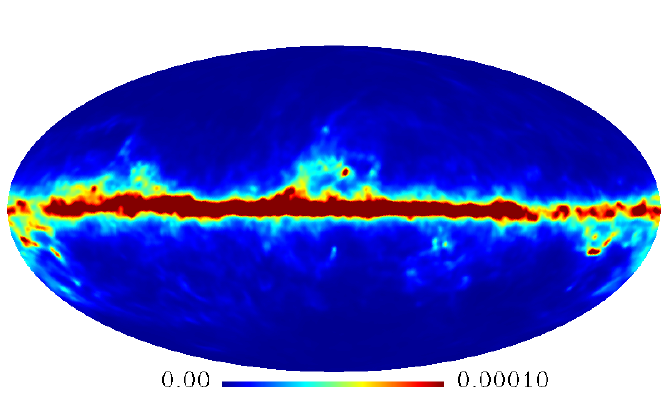}}}
\hbox{
\centerline{
\includegraphics[width=0.32\textwidth]{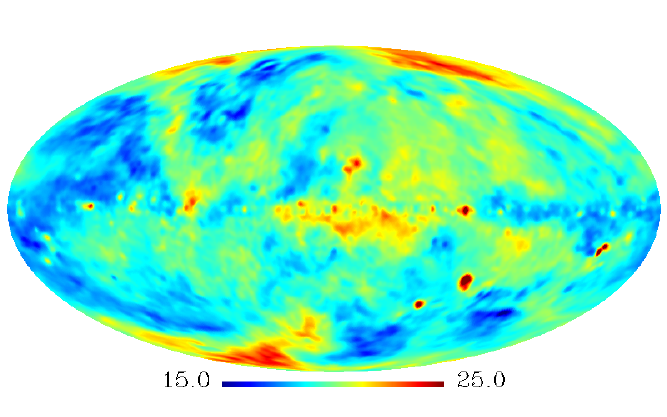}
\includegraphics[width=0.32\textwidth]{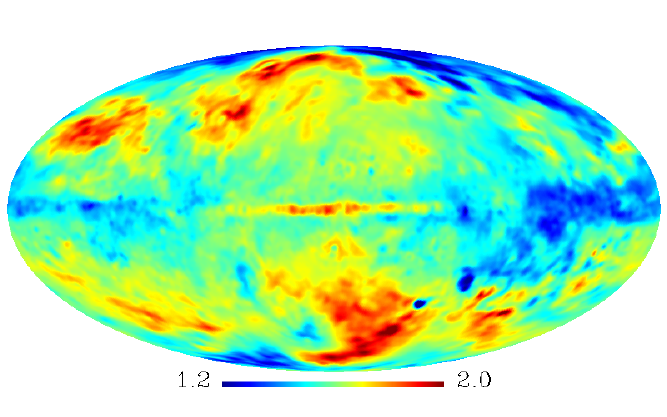}
\includegraphics[width=0.32\textwidth]{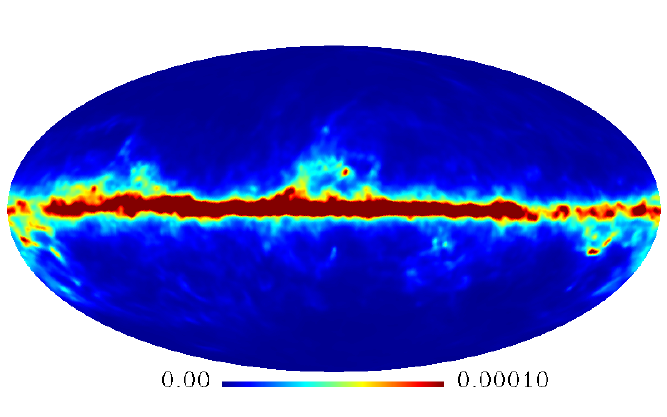}}}
\hbox{
\centerline{
\includegraphics[width=0.32\textwidth]{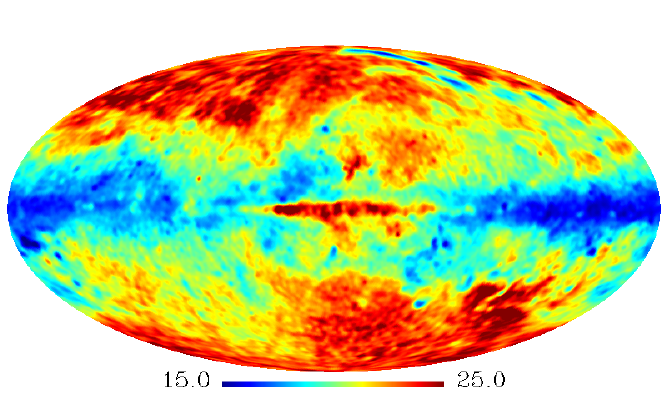}
\includegraphics[width=0.32\textwidth]{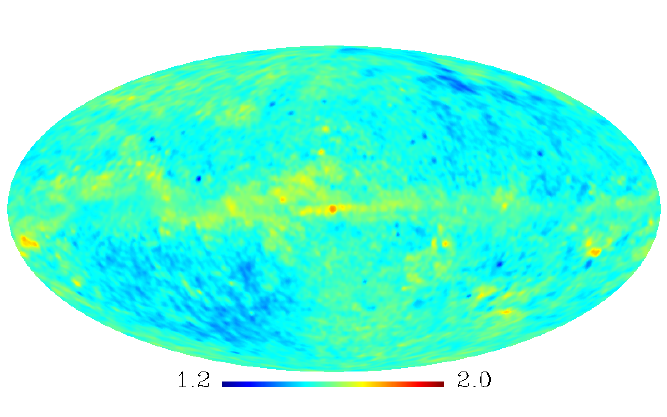}
\includegraphics[width=0.32\textwidth]{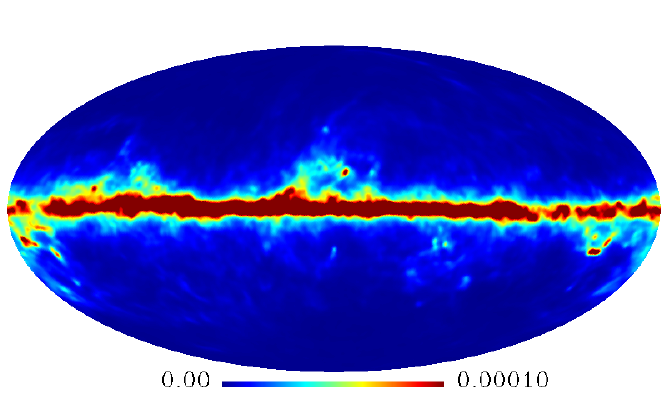}}}
\caption{The thermal dust emission parameter maps for $T_d$, $\beta$, and $\tau_{353}$ (from left to right), given by P13, P16, and C15 (from top to bottom). Note, that the optical depth of C15 is provided at a reference frequency of 545\,GHz, to allow for comparison we rescale down to 353\,GHz given the map of the spectral index.}
\label{fig2}
\end{figure*}

\begin{figure}[!htb]
\centering
\hbox{
\centerline{\includegraphics[width=0.5\textwidth]{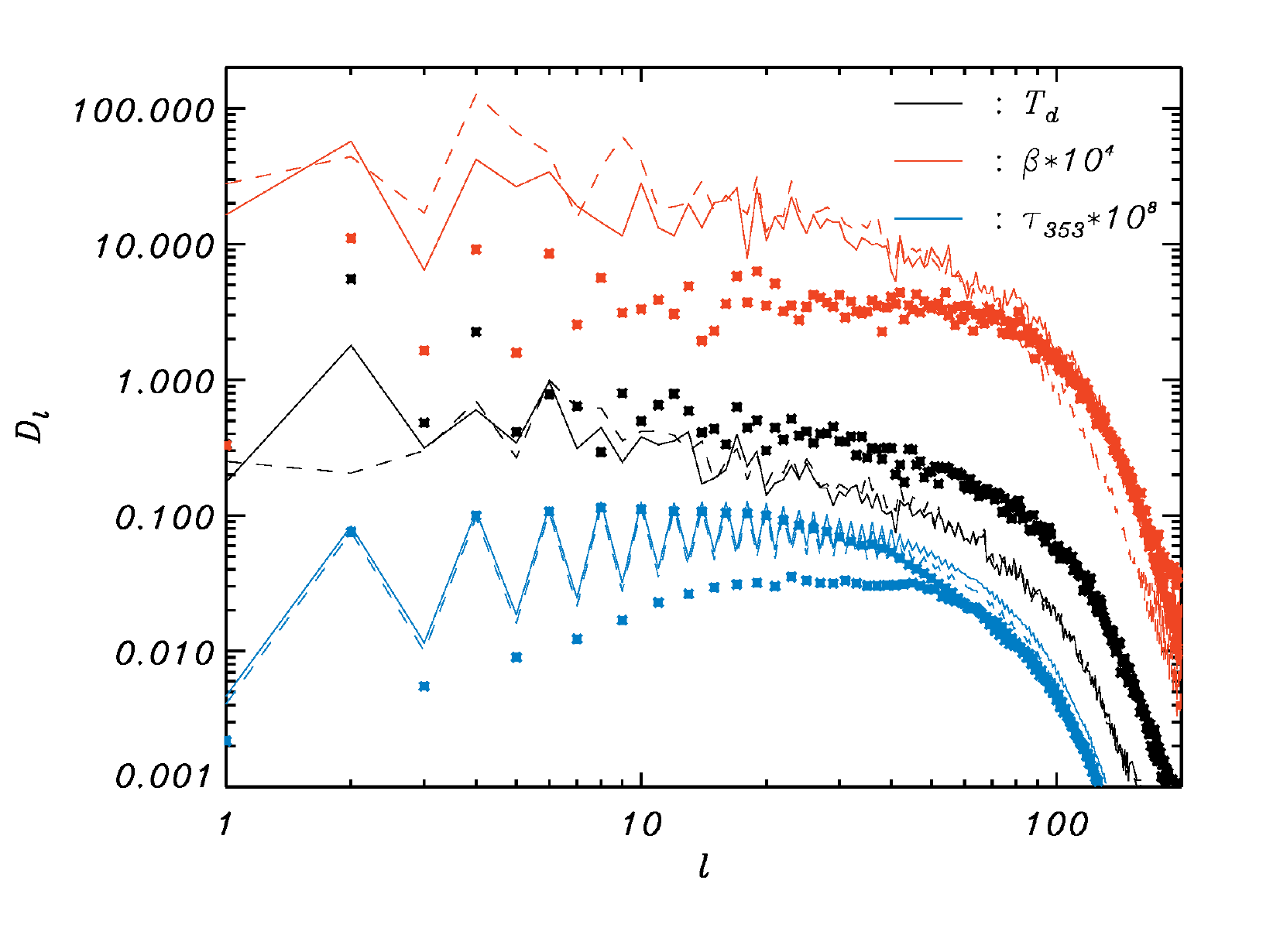}}}
\caption{Similar to fig.~\ref{fig1}, now with the power spectra of the C15 parameter maps (crosses). The cut-off at $\ell\gtrsim50$ corresponds to the $2^\circ$ Gaussian smoothing.}
\label{fig3}
\end{figure}

As justified in the last section, we now compare the parameter maps of P13, P16 and C15 after having smoothed them with a $\Theta=2^\circ$ Gaussian beam, see in fig.~\ref{fig2}. Even with the naked eye one notices that while the maps for the optical depth are very similar, both the spectral index and the temperature maps have strong differences, despite their similar mean values. We begin to analyze these differences with the power spectra of the maps, using the above definition, eq.~\ref{eq4}, see fig.~\ref{fig3}.

In comparison to P13 and P16, the C15 solution is characterized by a lower and flatter power spectrum for the spectral index $\beta$ at low multipoles, $\ell\lesssim70$, where $D(\ell)\propto const$. This seems to be compensated by more power in the temperature map on all scales. The optical depth experiences a loss of power towards higher $\ell$ at around $\ell\gtrsim40$. Needless to say, all the maps presented in fig.~\ref{fig2} reveal significant statistical anisotropy, clearly seen in the even-parity asymmetry of the power spectra (the amplitude of $D(\ell)$ for $\ell=$``even" is systematically greater then for $\ell=$``odd"), most pronounced in the map/the power spectrum of the optical depth, due to the concentration of the signal along the Galactic plane -- a source of ``even"-power
dominance~\cite{kimnas1,kimnas2,kimnas3}, which is most pronounced in the Commander maps.

\begin{figure*}[!htb]
\hbox{
\centerline{
\includegraphics[width=0.32\textwidth]{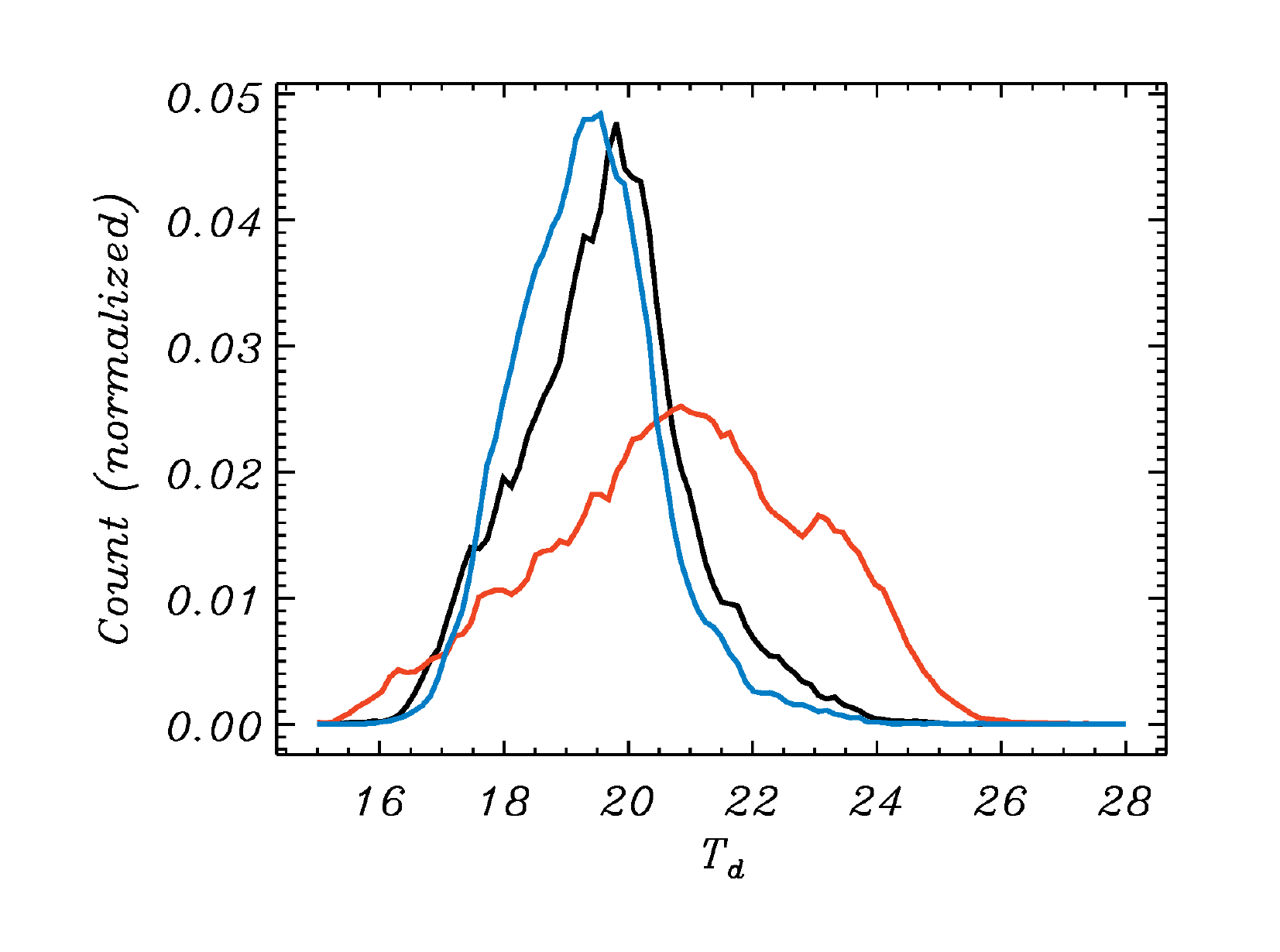}
\includegraphics[width=0.32\textwidth]{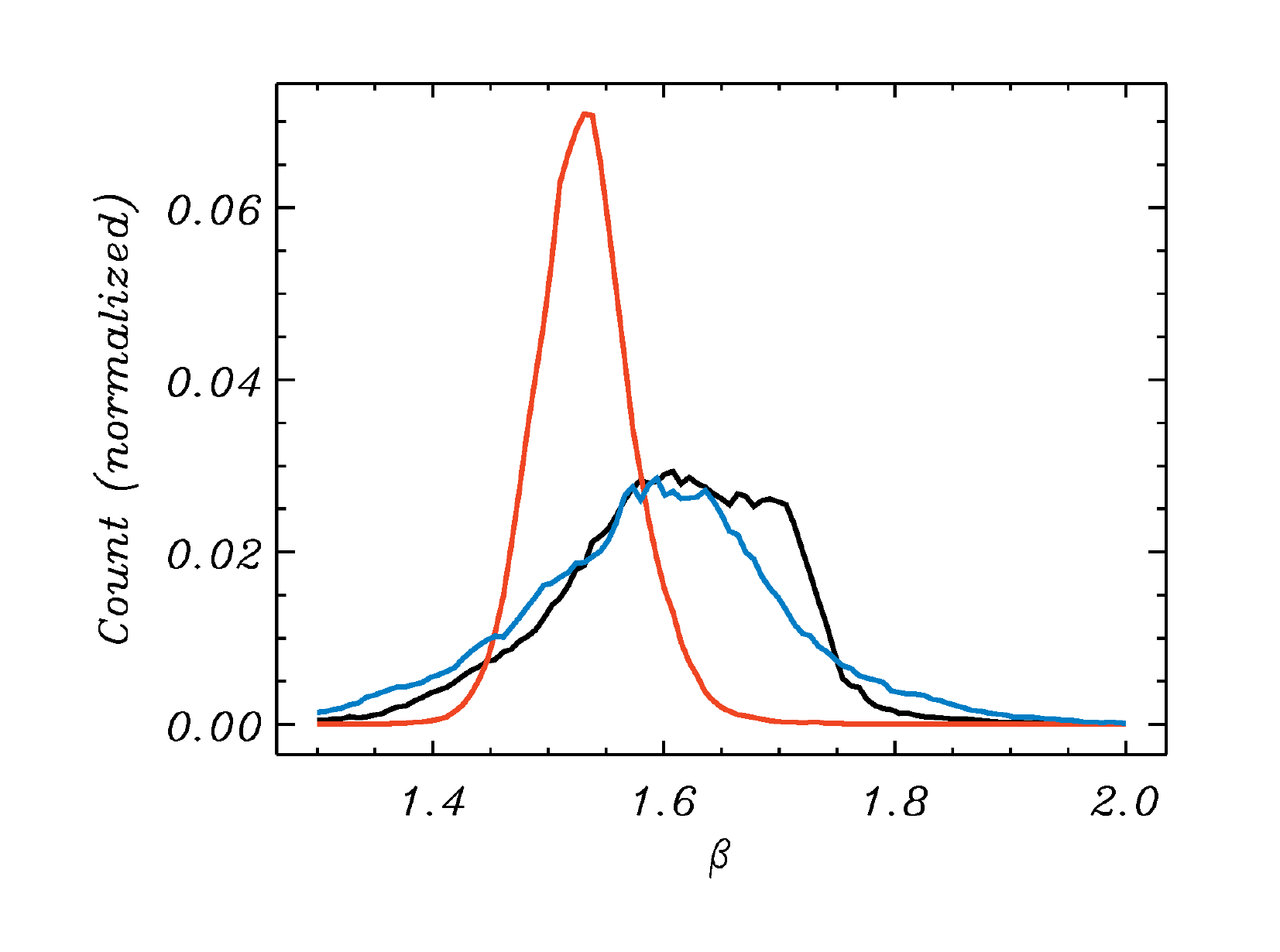}
\includegraphics[width=0.32\textwidth]{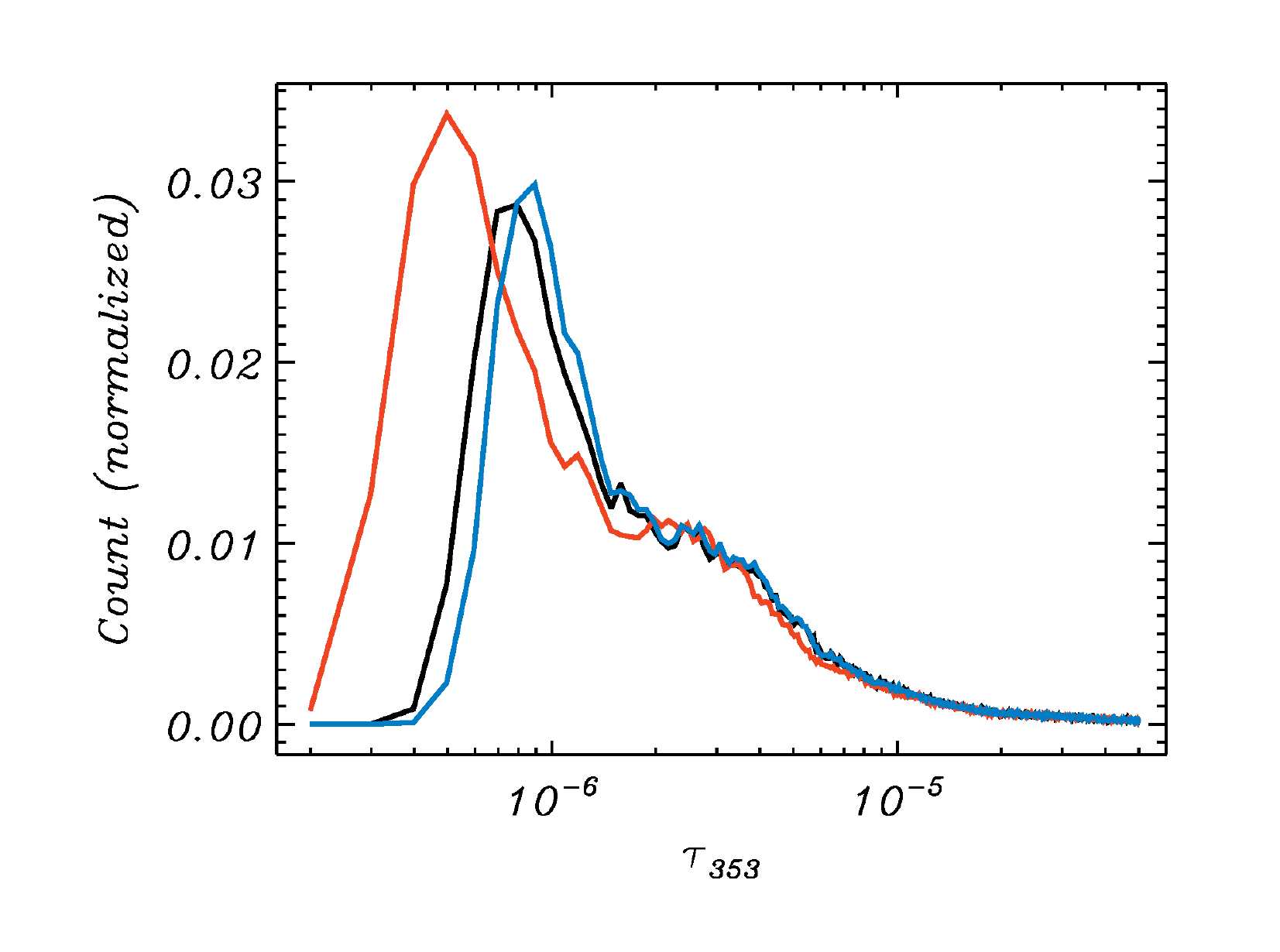}}}
\caption{The  distribution functions for $T_d$, $\beta$, and $\tau_{353}$ (from left to right), given by P13 (black), P16 (blue) and C15 (red)for the full sky maps.
}
\label{fig4}
\end{figure*}

We proceed to compare number densities, $P(X)$, of all the parameter maps(at a \textsc{HEALPix} resolution of $N_{\text{Side}}=128$) in the intervals $[X-\Delta X,X+\Delta X]$ with $\Delta X=10^{-2}X_{max}$, where $X_{max}$ is the maximum of the corresponding parameter, see fig.~\ref{fig4}. The distributions are normalized to the total number of pixels in the maps. It becomes obvious that the C15 distributions reflect precisely the choice of priors. The Gaussian priors in C15 were chosen to be $\beta^{C15}=1.55\pm 0.1$ and $T^{C15}=23\pm 3$; $\tau$ is calculated from a combination of the previous parameters and the dust amplitude, a parameter only constrained to be positive. Note that  in P13 the allowed ranges of variation are $1.0\le\beta^{P13}\le 2.5$ and $10\le T^{P13}\le 60$ K.
In spite of different treatment regarding the CIB, the temperature and optical depth distributions of P13 and P16 look very similar. Only the distributions of the spectral index exhibit a discrepancy at $\beta\simeq1.7$; it stands to reason that this difference is likely caused by the CIB.
\begin{figure}[!htb]
\hbox{
\centerline{
\includegraphics[width=0.45\textwidth]{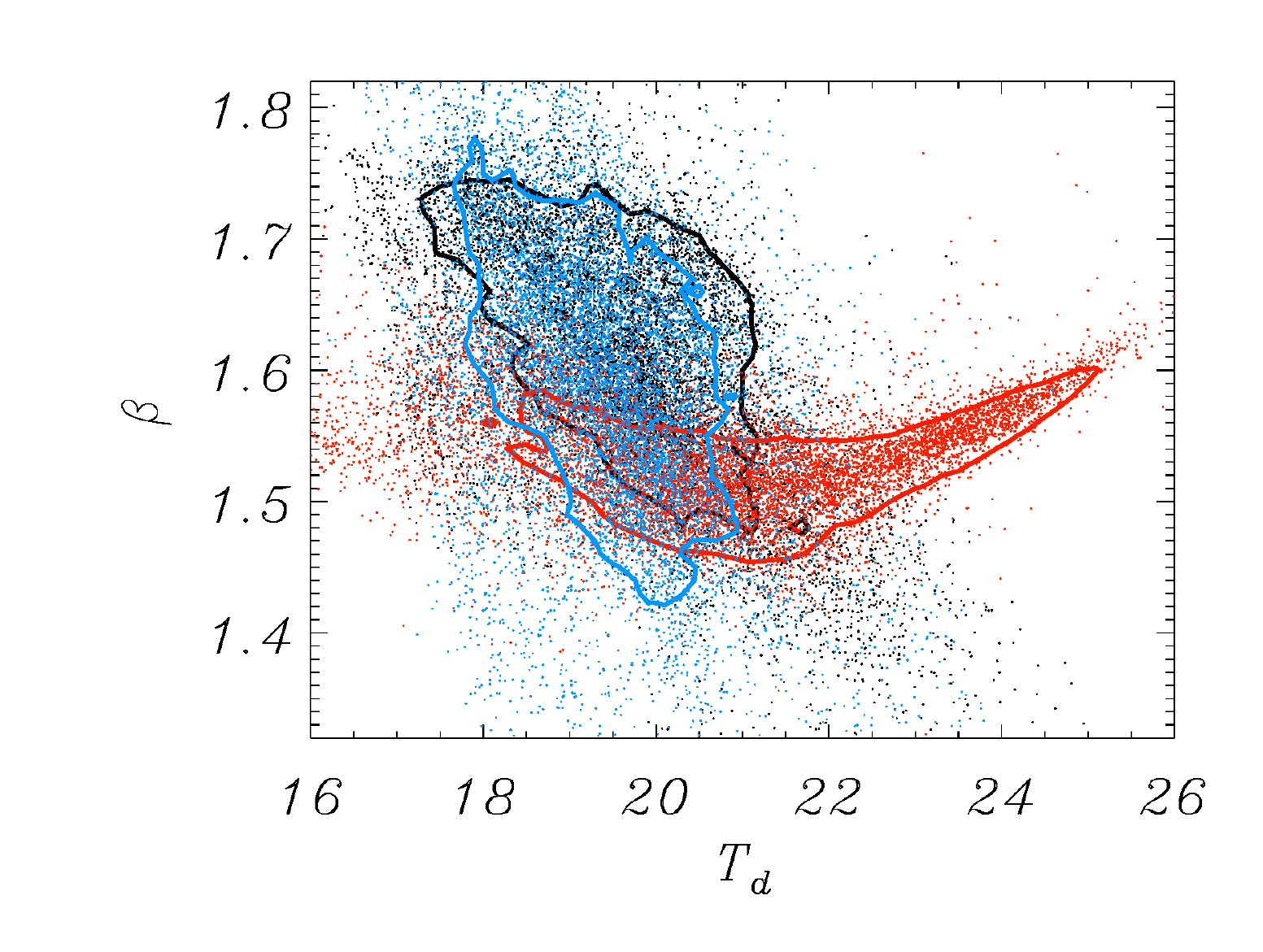}}}
\hbox{
\centerline{
\includegraphics[width=0.45\textwidth]{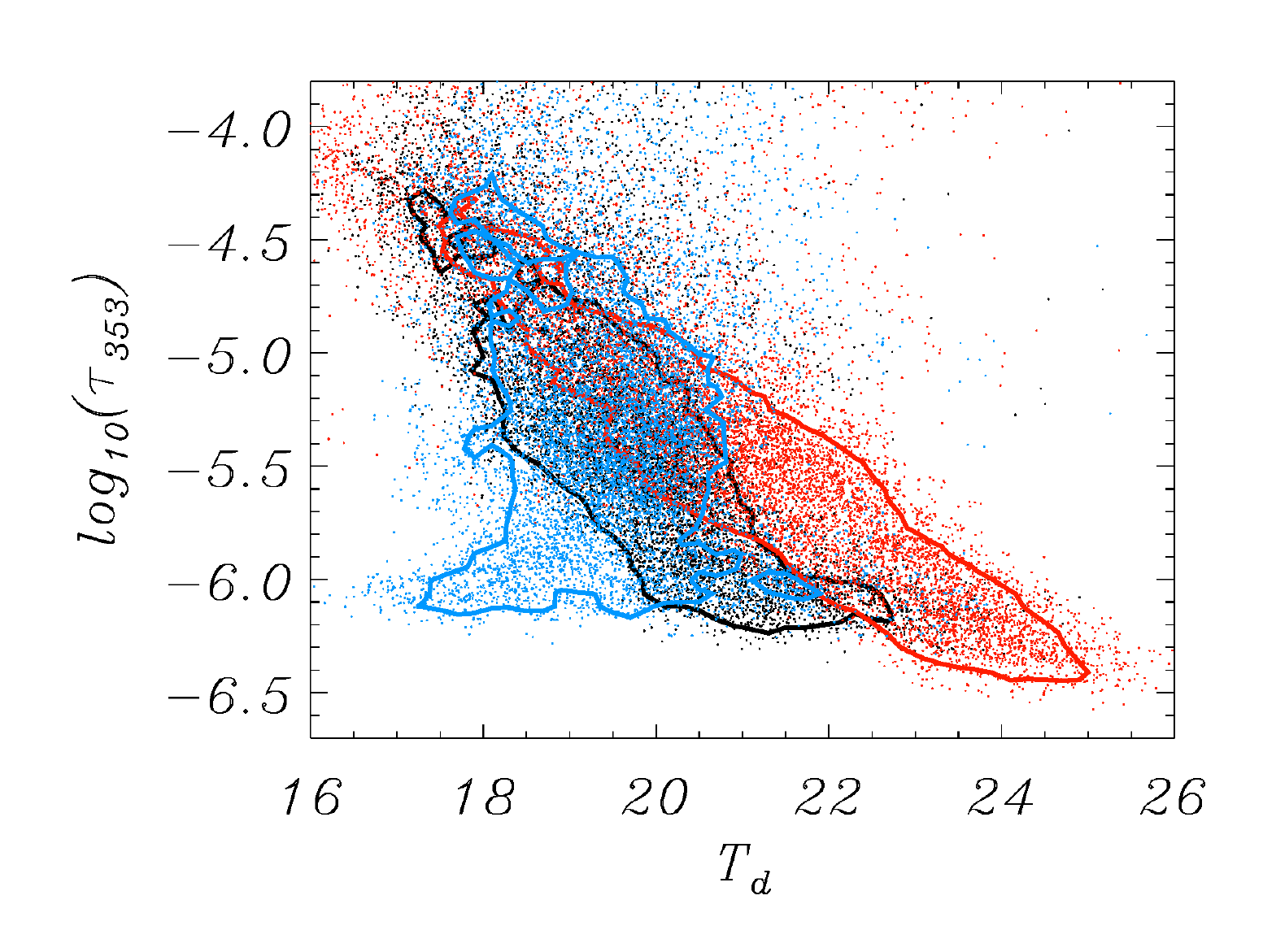}}}
\caption{T-T plots for $T_d$ versus $\beta$ (top), and for $\tau_{353}$ versus $T_d$ (bottom), of P13 (black), P16 (blue) and C15 (red). The solid contours correspond to the $68\%$ confidential level of the 2-dimensional distributions of the points.}
\label{fig5}
\end{figure}

Finally, we compute full-sky cross-correlations between $T_d$ and $\beta$ as well as $T_d$ and $\tau$, for each of the solutions, see fig.~\ref{fig5}. The negative correlation between $T_d$ and $\beta$ is well known from P13~\cite{PlanckDust03} and is also reproduced by P16. Also from the P13 and P16 maps in fig.~\ref{fig5} we find that for most regions of the sky lower temperature regions are accompanied by higher spectral index regions and vice versa. While the C15 solution at low temperatures shows negative correlation as well -- yet less pronounced -- it turns to become strongly positive at temperatures $T_d\gtrsim21\,K$. Again referring to fig.~\ref{fig2} we find that the regions in which dust temperatures this high occur, happen to lie at higher Galactic latitudes $|b|\gtrsim50^\circ$, those regions where the CIB begins to dominate. The negative correlation between $T_d$ and $\tau$ present for the P13 solution, is well reproduced by C15, however marginally for P16.

\section{Local comparison of the solutions}
\label{sec:local}

\begin{figure}[!htb]
\hbox{
\centerline{
\includegraphics[width=0.5\textwidth]{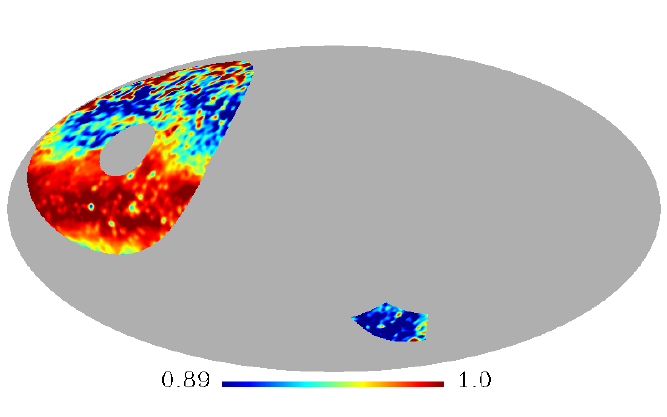}}}
\caption{$I_{353}^{P16}/I_{353}^{P13}$ for the BICEP2 zone (lower right), and the NCP zone (upper left). }
\label{fig6}
\end{figure}

\begin{figure}[!htb]
\includegraphics[width=0.23\textwidth]{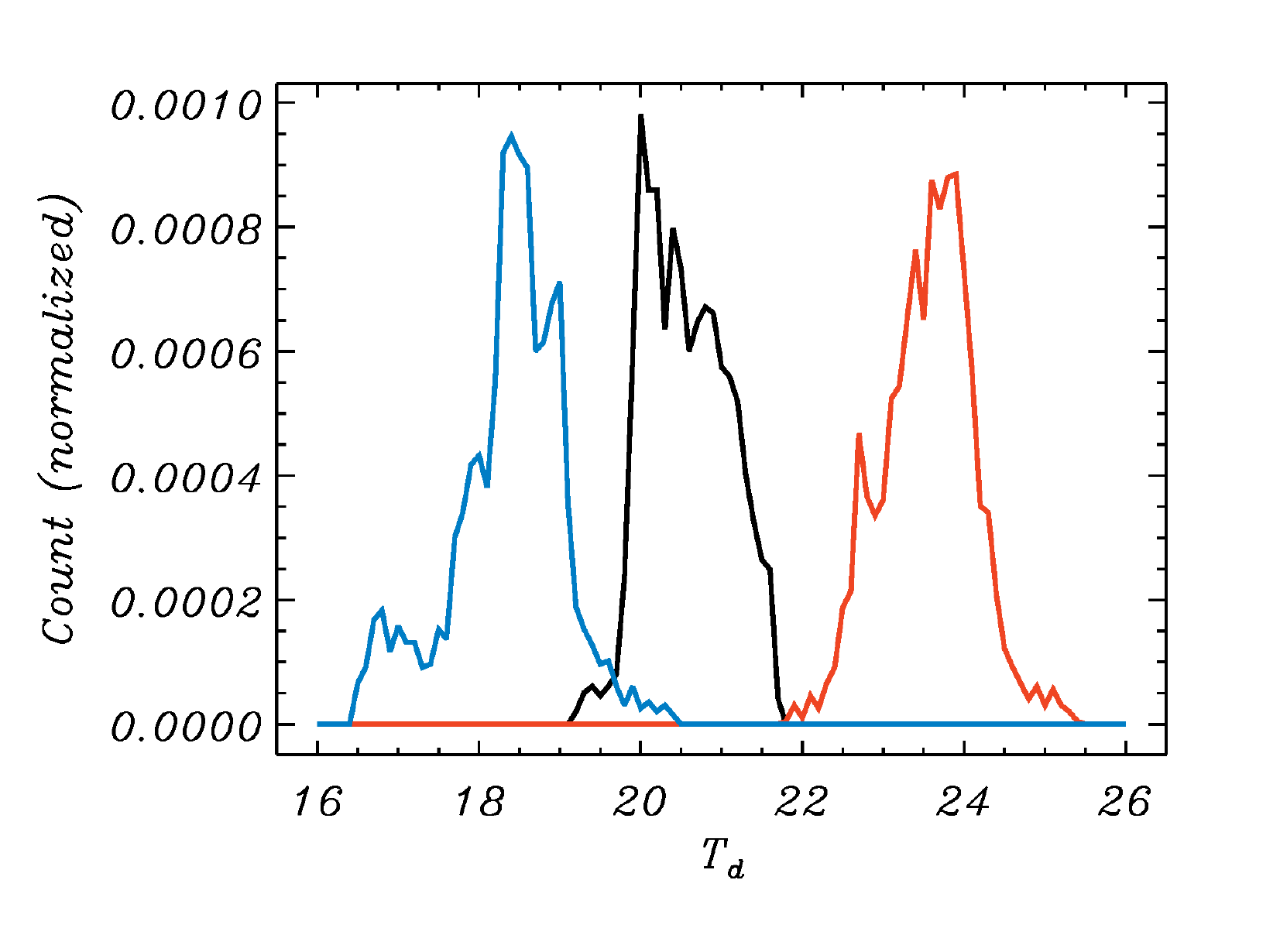}
\includegraphics[width=0.23\textwidth]{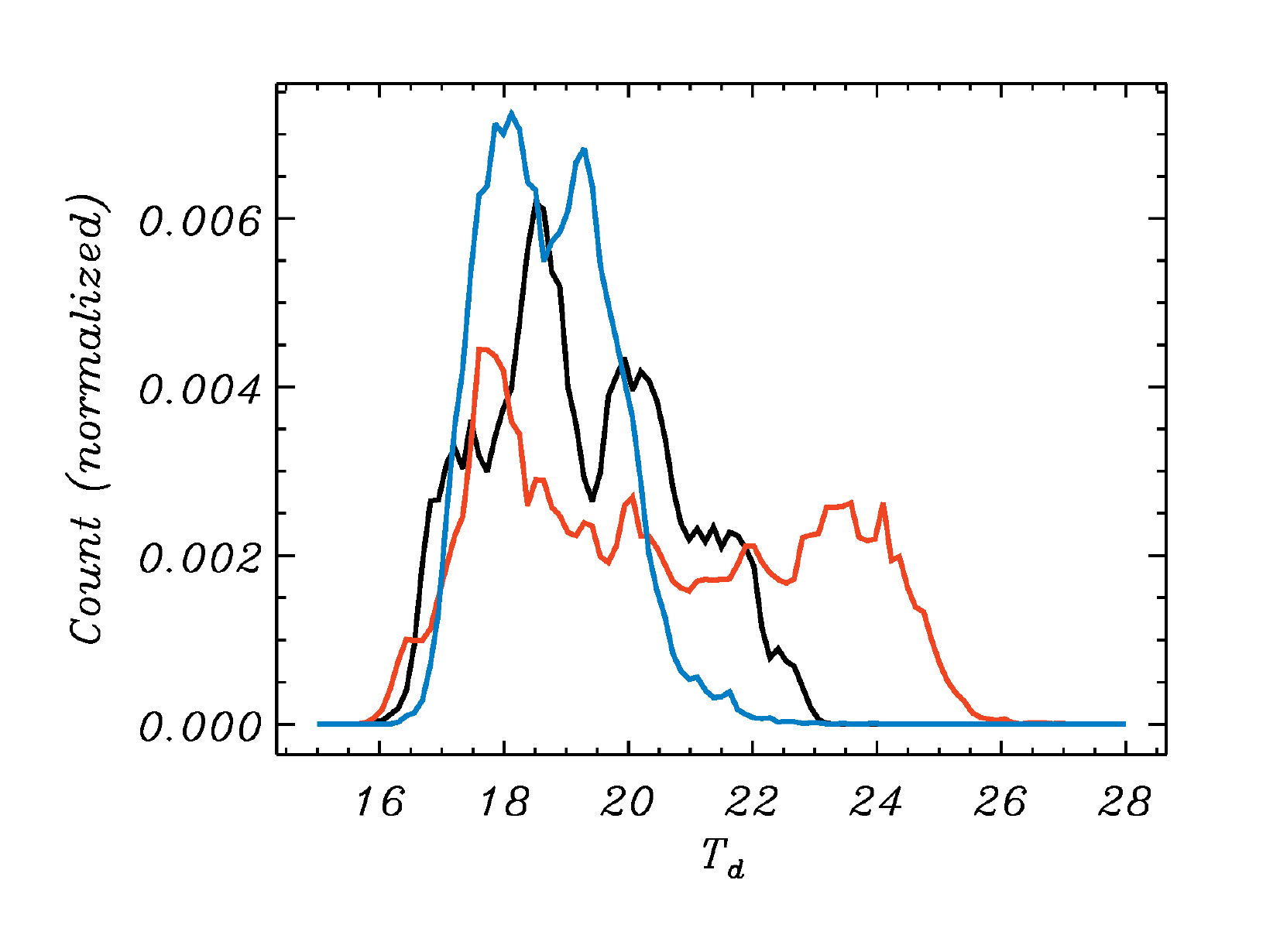}  \\
\includegraphics[width=0.23\textwidth]{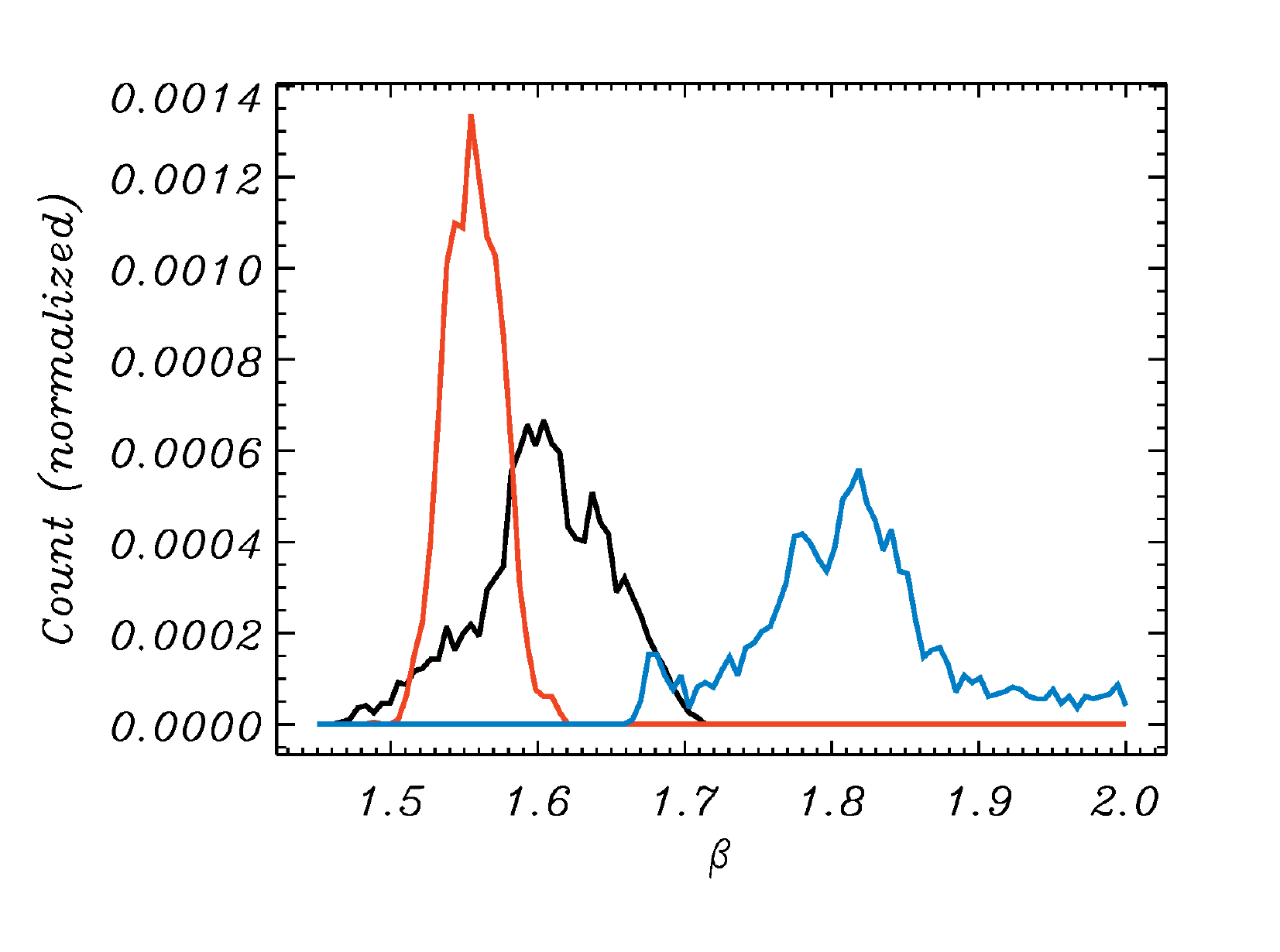}
\includegraphics[width=0.23\textwidth]{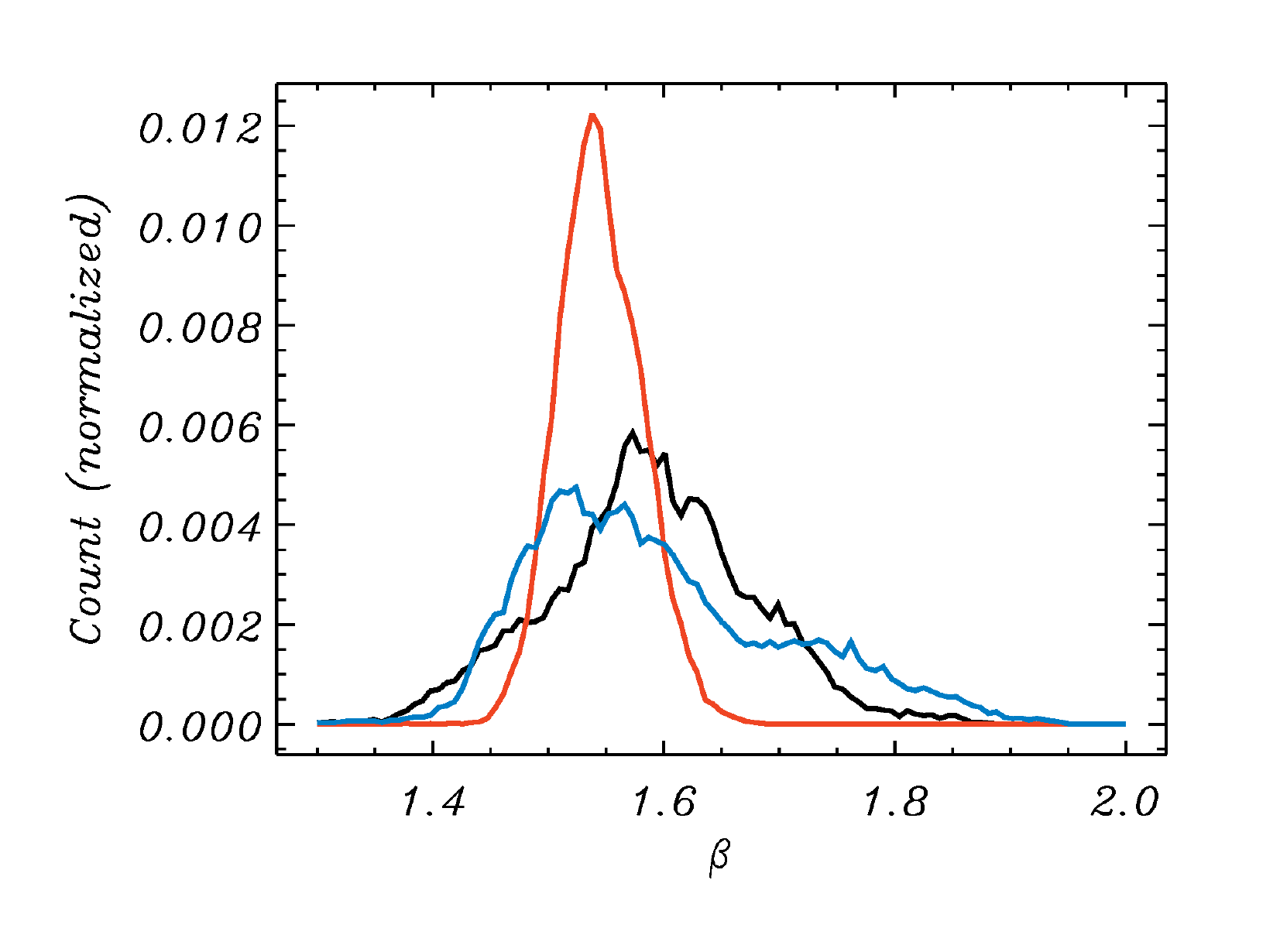}  \\
\includegraphics[width=0.23\textwidth]{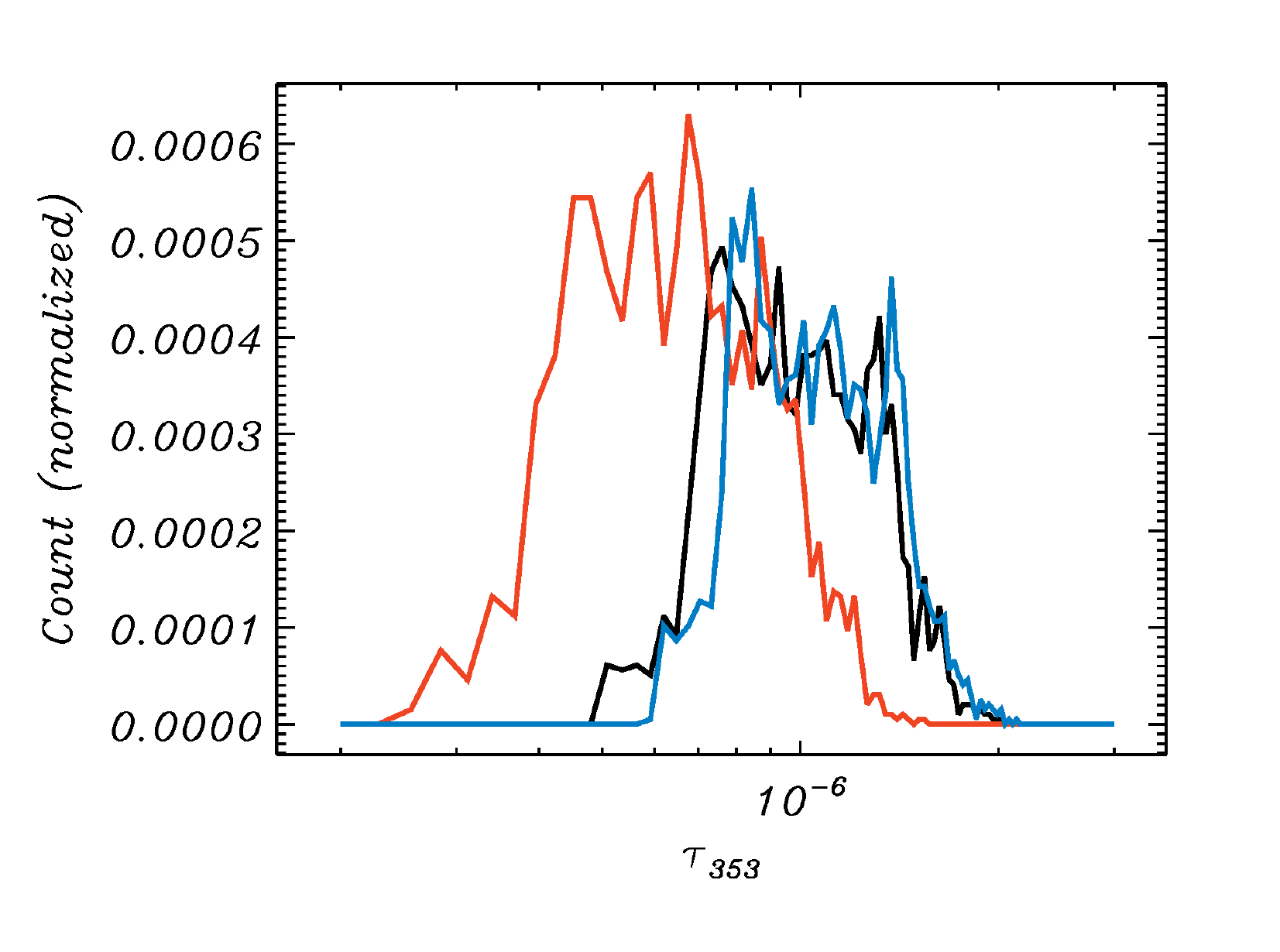}
\includegraphics[width=0.23\textwidth]{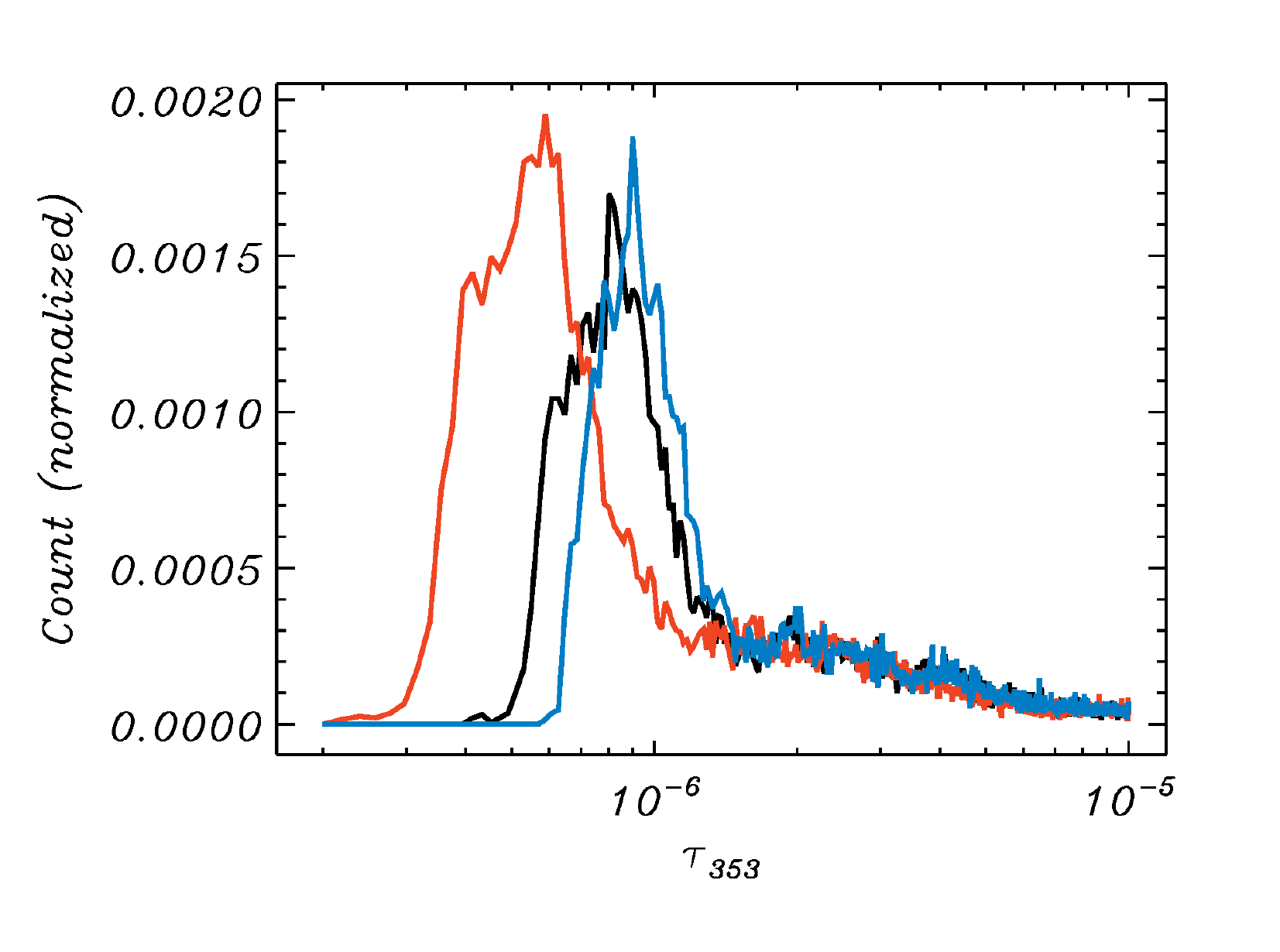}
\caption{The  distribution functions for the dust temperature $T_d$, the spectral index  $\beta$, and the optical depth $\tau_{353}$ (from top to bottom), given by P13 (black), P16 (blue) and C15 (red), restricted to the  BICEP2 zone (left), and the NCP zone (right).
}
\label{fig7}
\end{figure}

In this section we would like to underline that the variability of the dust emission over the sky and the difference of the solutions forbids a flat-rate use of one particular choice of dust parameters to describe different regions on the sky properly. For this purpose we pick two regions on the sky as examples for areas under investigation in ground-based CMB experiments: the BICEP2 region in the Southern hemisphere, and a second region around the North celestial pole (NCP), observed by, e.g. QUIJOTE~\cite{QUIJOTE1,QUIJOTE2,QUIJOTE3,QUIJOTE4,QUIJOTE5,QUIJOTE6} or DeepSpace\cite{deepspace}, see fig.~\ref{fig6}, where we show the ratio of the dust emission intensity in P16 to that in P13. While in the BICEP2 zone the intensities differ by a comparably constant offset, the NCP region shows a strong gradient from the Galactic plane, were the dust models agree fairly well, decreasing to higher latitudes, however, with many small features. Again, we investigate the distributions of the parameters of the different solutions, now separately for the two regions, see fig.~\ref{fig7}. The results differ considerably. Especially the dust temperatures and the spectral indices in the BICEP2 zone disagree strongly. As the latter are used for the extrapolation of dust intensities from higher to lower, CMB-friendly frequencies, this difference will immediately correspond to a difference in the obtained dust contamination.

For future evaluation of data by planned CMB experiments, which due to their frequency coverage cannot perform independent dust emission estimates, it is crucial to find concordance in which dust parameters to use. For frequencies below $\nu_0=353\,$GHz and characteristic dust temperatures, eq.~\ref{eq1} reduces to
\begin{eqnarray}
I_{\nu}(\textbf{n}) \propto \tau(\textbf{n})T_d(\textbf{n})\left(\frac{\nu}{\nu_0}\right)^{2+\beta(\textbf{n})}
\label{eqRJ}
\end{eqnarray}
The parameters in the BICEP2 zone follow $\beta^{C15}<\beta^{P16}$, $T_{d}^{C15}>T_{d}^{P16}$, and $\tau^{C15}\simeq\tau^{P16}$. According to eq.~\ref{eqRJ} the intensity of dust emission at $\nu_0<353\,$GHz would be systematically greater in the C15 solution than in P16.

In the NCP region the distribution of the thermal dust intensity is more complicated, as there is much more variation present than in the BICEP2 zone. From the number densities only, this is most apparent in the broad distribution of dust temperatures peaked at 17 and 23\,K, corresponding to the low and the high Galactic latitude parts of the region.

The comparison of these two zones clearly illustrates that the evaluation of the dust emission intensity at different frequencies cannot be done independently of the region investigated via e.g. the use of average values for the dust parameters. The 5-20\% variations of these parameters will generate artificial signals, interfering with any measurement of the $B$-mode of polarization.

\begin{figure}[!htb]
\includegraphics[width=0.23\textwidth]{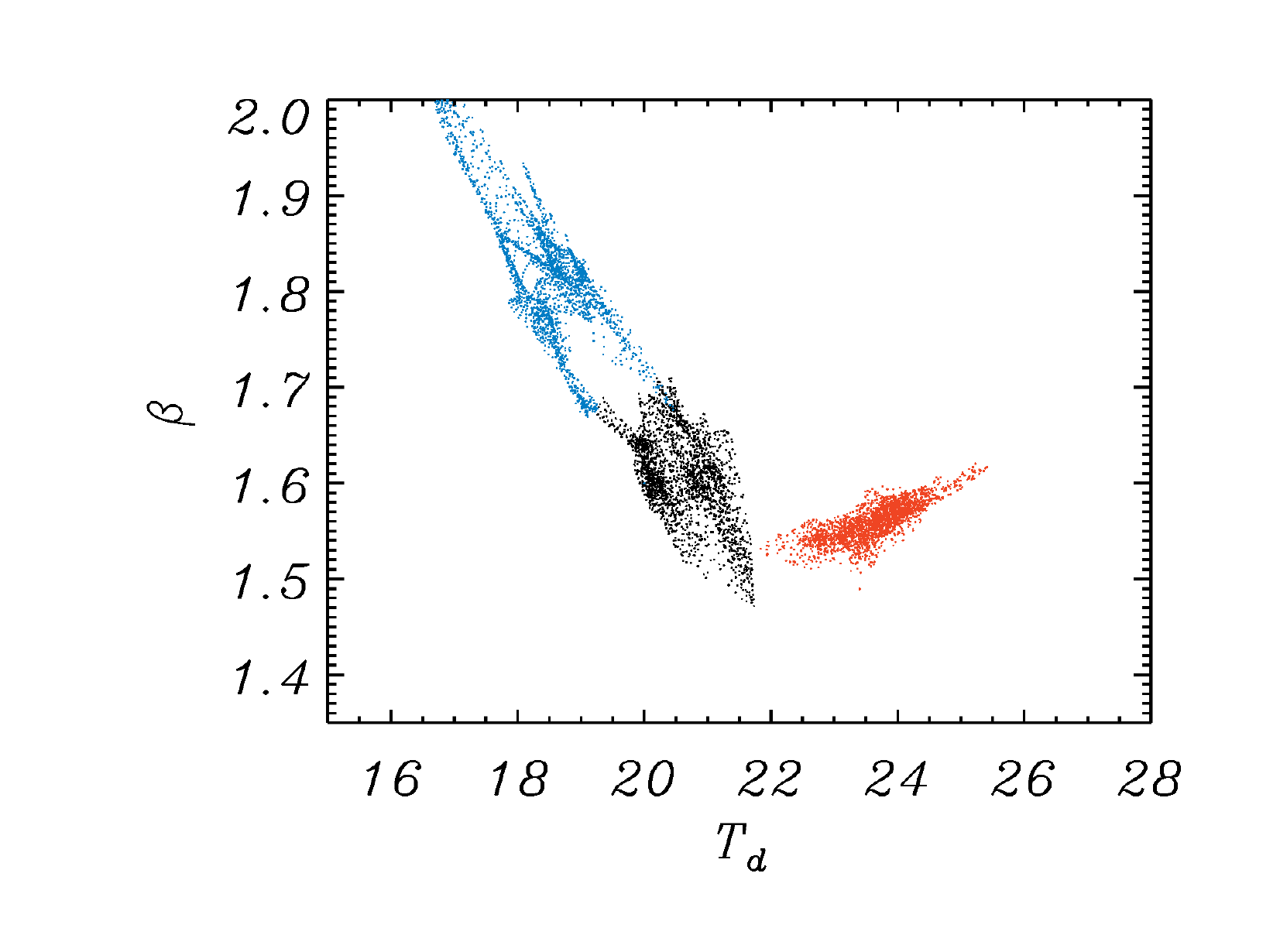}
\includegraphics[width=0.23\textwidth]{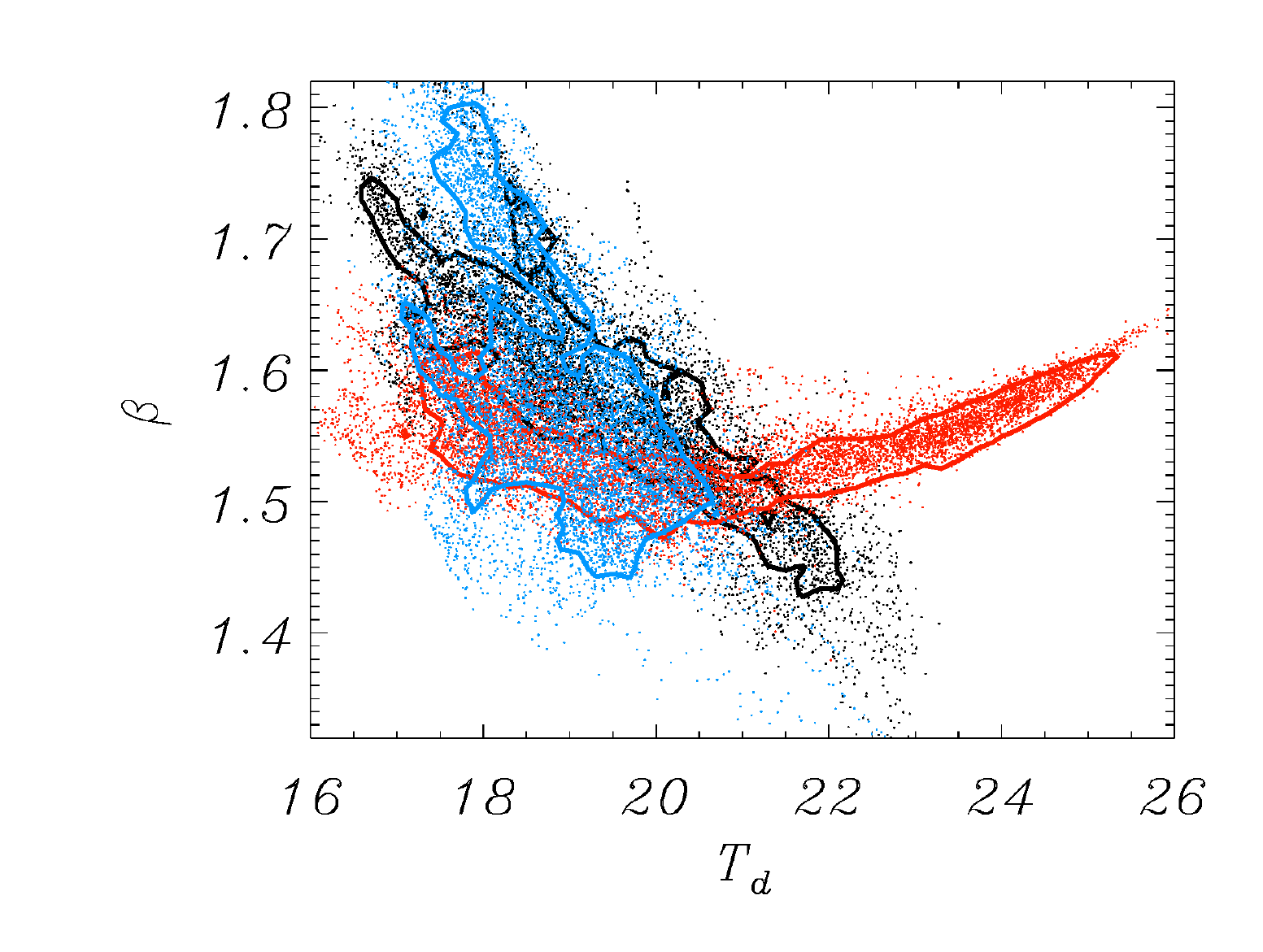}  \\
\includegraphics[width=0.23\textwidth]{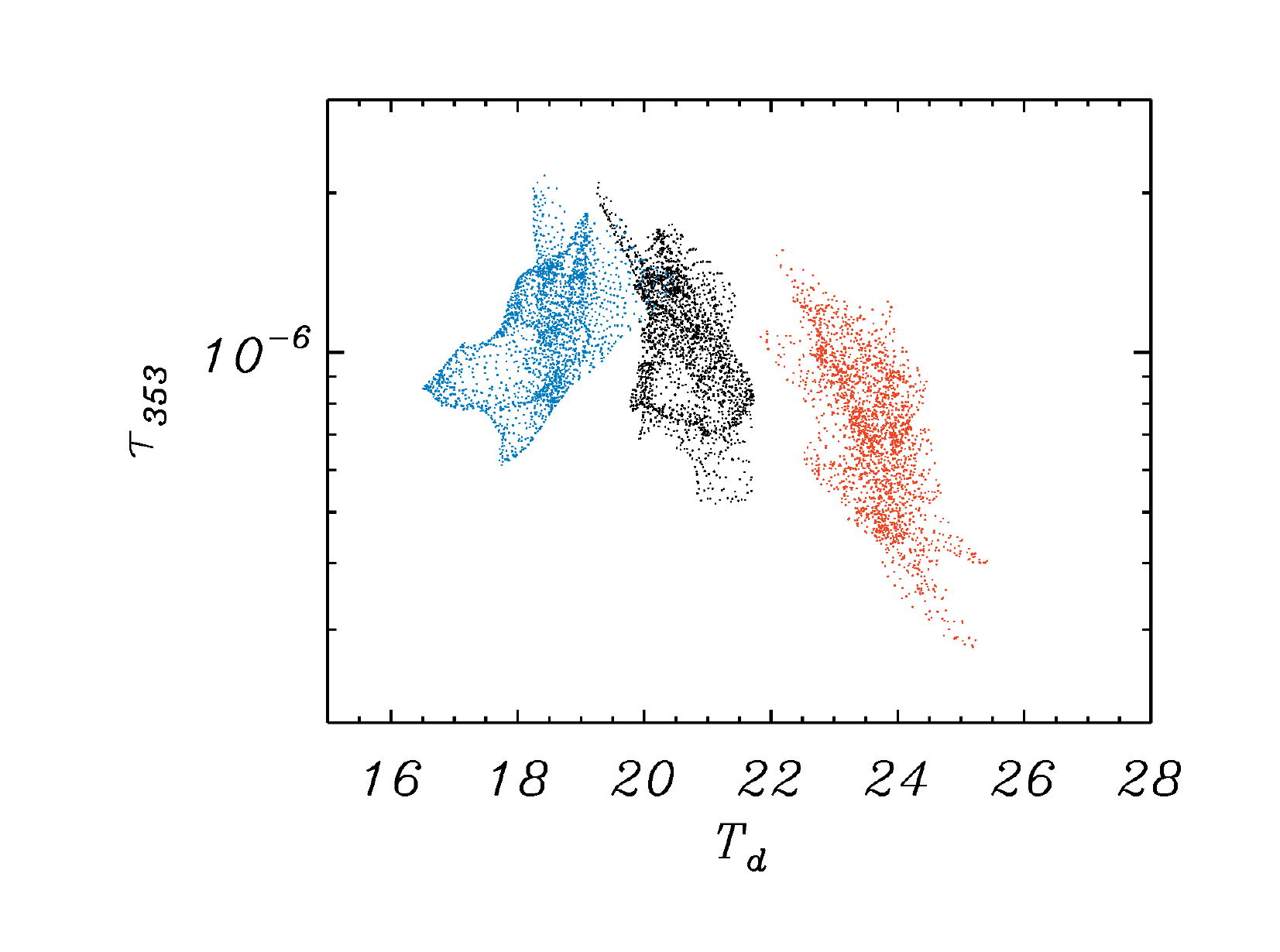}
\includegraphics[width=0.23\textwidth]{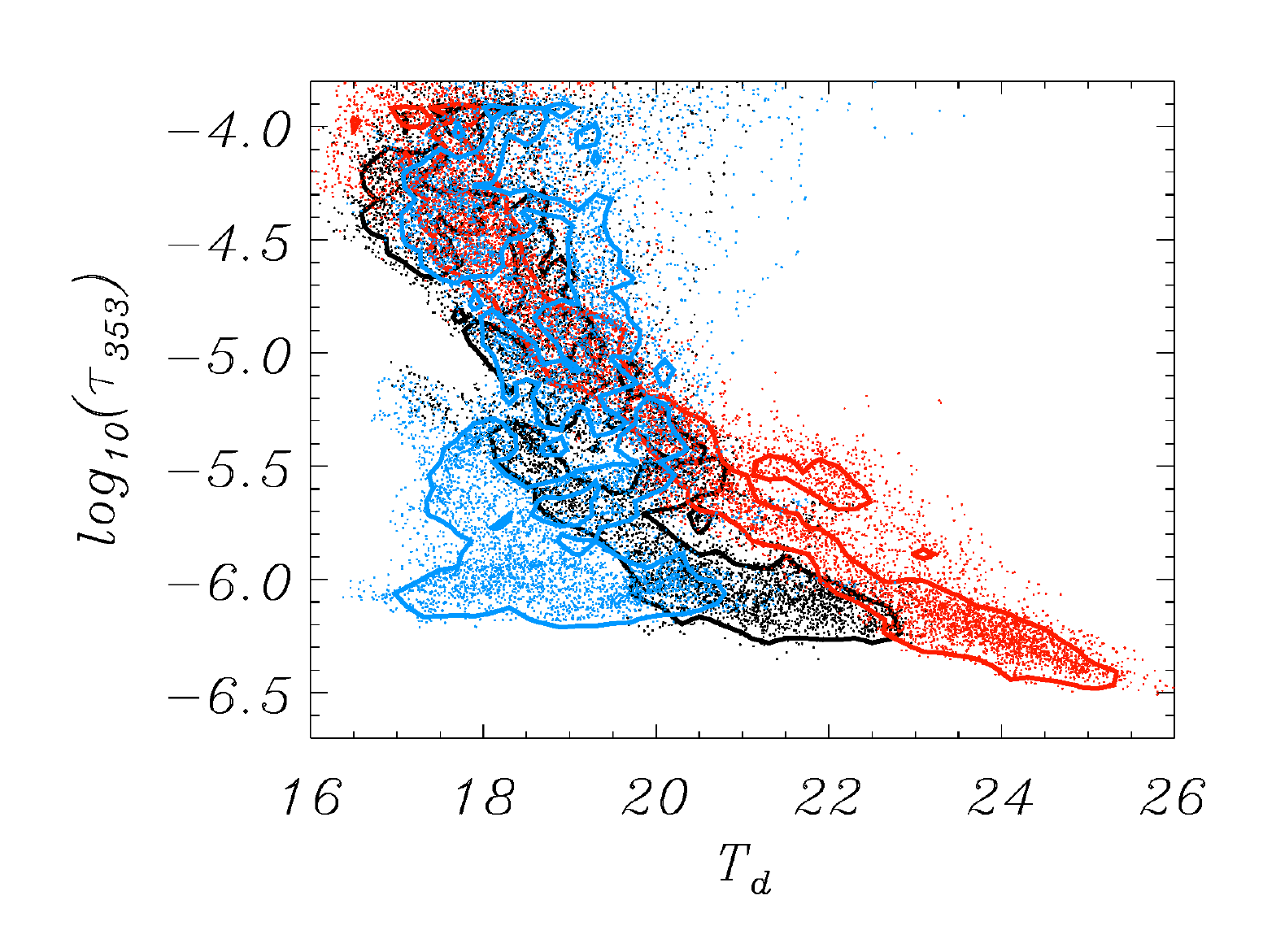}  \\
\caption{Similar to fig.~\ref{fig5}, now restricted to the BICEP2 zone (left) and the NCP zone (right). Note, that since the BICEP2 zone contains far less pixels, we choose not to plot contour lines.}
\label{fig8}
\end{figure}

For completeness, we also show the $T_d$-to-$\beta$ and $T_d$-to-$\tau$ correlations in fig.~\ref{fig8} for both zones in question.

\section{Discussion and Conclusion}
\label{sec:sum}

In this paper, we compared the three full-sky solutions to thermal dust emission, P13, C15, and P16 from the Planck Collaboration and found significant variations in their respective parameters, the spectral index $\beta$, the optical depth $\tau$ and the dust temperature $T_d$. Thereby P16 is the only solution which was cleaned of the CIB. At high resolution, the difference to P13 is distinctly visible in their power spectra at $\ell\gtrsim50$.
\begin{table}[!htb]
\caption{Cross-correlation coefficients between $T_{d}$ and $\beta$ of the P13, P16 and C15 solutions for the full sky.
}
\begin{tabular}{|l|c|c|c|c|c|c|} \hline
& $T_{d,P13}$ & $T_{d,P15}$ & $T_{d,P16}$ & $\beta_{P13}$ & $\beta_{C15}$ & $\beta_{P16}$ \\ \hline
$T_{d,P13}$ &    1.00    &    0.71 &    0.67 &   -0.60 &   -0.16 &   -0.06\\ \hline
$T_{d,C15}$ &        &   1.00   &    0.19 &   -0.11 &   -0.04 &    0.52\\ \hline
$T_{d,P16}$ &       &         &   1.00   &   -0.30 &   -0.22 &   -0.51\\ \hline
$\beta_{P13}$ &       &        &        &   1.00   &    0.27 &    0.51\\ \hline
$\beta_{C15}$ &       &        &        &         &    1.00  &    0.21\\ \hline
$\beta_{P16}$ &       &        &        &         &         &   1.00  \\ \hline
\end{tabular}
\label{tab:cc_td_beta}
\end{table}

\begin{table}[!htb]
\caption{Cross-correlation coefficients between $T_{d}$ and $\tau_{353}$ for the P13, P16 and C15 solutions. Note that here we use a ring mask to exclude $\pm10^\circ$ around the Galactic  plane.
}
\begin{tabular}{|l|c|c|c|c|c|c|} \hline
& $T_{d,P13}$ & $T_{d,C15}$ & $T_{d,P16}$ & $\tau_{353,P13}$ & $\tau_{353,C15}$ & $\tau_{353,P16}$ \\ \hline
$T_{d,P13}$ &    1.00 &    0.64 &    0.66 &   -0.57 &   -0.55 &   -0.59\\ \hline
$T_{d,C15}$ &         &    1.00 &    0.07 &   -0.62 &   -0.63 &   -0.64\\ \hline
$T_{d,P16}$ &         &         &    1.00 &   -0.20 &   -0.18 &   -0.21\\ \hline
$\tau_{353,P13}$&         &         &         &    1.00 &    1.00 &    1.00\\ \hline
$\tau_{353,C15}$&         &         &         &         &    1.00 &    0.99\\ \hline
$\tau_{353,P16}$&         &         &         &         &         &    1.00\\ \hline
\end{tabular}
\label{tab:cc_td_tau}
\end{table}
After having smoothed the maps, comparison with those of C15 shows a strong deviation of the distribution of parameters to those of both former ones, obviously caused by the priors imposed in the C15 method. Hints towards an influence of the CIB also in the C15 maps was found by the systematically higher dust temperature in regions of higher Galactic latitude $|b|\gtrsim50^\circ$, which is positively correlated with the spectral index, opposite to the case at lower temperatures/lower Galactic latitudes.
However, no upturn was found for the $T_d-\beta$ correlation of P13. We summarize all cross-correlations between $T_d$ and $\beta$, and $T_d$ and $\tau$ in Table I and Table II, where for the sake of completeness we also include the correlations across methods. Note, that the $T_d-\beta$ anti-correlations are significant only for the P13 and P16 models.
\begin{table}
\caption{List of the monopole correction for the 357-3000 GHz maps by P13 and P16, in MJy/sr.}
\centering
\begin{tabular}{|c|c|c|c|c|}
\hline
  $\nu$ [GHz]      &     353 &  545 &     857 &     3000 \\ \hline
P13     & 0.085   & 0.095  & 0.093 & -0.174 \\ \hline
P16     & 0.125 &  0.336 &  0.556 &  0.113 \\ \hline
\end{tabular}
\label{tab:tbl2}
\end{table}
Also note, that the parameters of the P13 and P16 models critically depend on the monopole corrections, presented in Table 3. Potentially  these  corrections could be one of the major sources of uncertainties for future CMB $B$-mode experiments.

We have also investigated the distributions of the spectral index $\beta$, the optical depth $\tau$ and the dust temperature $T_d$ maps in all three models in two regions of the sky: the BICEP2 zone and one around the North Celestial Pole. We have shown that in the BICEP2 zone the spectral index for P16 model is shifted to values around $\beta\simeq 1.8$ compared to the C15 spectral index at $\beta\simeq 1.55$. In the NCP zone the behavior of the spectral indices and the dust temperatures are very different than those in the BICEP2 zone. Here C15 predict the temperature distribution function with two maxima at 17K and 23K. Most likely, the highest values of the dust temperature in NCP zone are related to the CIB contamination (mainly due to its monopole) of the C15 MBB parameters.

In conclusion, all methods compared here agree to about $5-20\%$. However, in the light of current ambitions of planned $B$-mode experiments, much higher accuracy is needed. It is curious to note, that the different solutions agree best along the Galactic plane at $|b|\lesssim50^\circ$--those regions which are usually excluded from analyses due to the large dust emission amplitudes. The regions with $|b|\gtrsim50^\circ$ show less emission by the Galaxy, however the increased relative contribution of CIB, instrumental noise possibly other components as anomalous microwave emission, in the context of the Commander model investigated in~\citep{Hau15}, causes disagreement among the different dust emission solutions. Due to rapid changes of the dust parameters over the sky, we warn against the use of average values of the model's parameters. The residuals introduced by this approach would themselves act as a ''new" foreground, leading to biases in the polarized component separation (Such biases are nicely shown in form of altered spectral shapes presented in~\citep{Chluba2017} for intensity measurements). Assuming that one knows the correct model(s) describing the foregrounds, supplemented by observations at high enough frequencies above 300~GHz, we strongly promote local (pixel- and multipole-based) fitting methods ensuring proper treatment of both thermal dust emission and CIB, despite its higher computational cost. Only then can one ensure the reliable extrapolation of foreground emission to the frequency range of interest, around $100\,$GHz, where ultimately also low frequency foregrounds must be treated to the same level of precision. In general, the next generation of $B$-mode missions needs more sophisticated methods of foreground removal, which carefully treat the averaging effects described here, as well as those along the line-of-sight, possibly based on further development of the Commander and P16 algorithms including the CIB and less strong priors on the parameters.

\ifx\usePRL\defined
\begin{acknowledgments}
\fi

\ifx\useJCAP\defined
\section{Acknowledgments}
\fi

We thank Carlo Burigana, Per Rex Christensen, Jacques Delabrouille, Hans Kristian Eriksen, Andrew Jackson, and Philipp Mertsch for comments on the draft and useful discussions. We also thank the anonymous referee for useful comments on improving the manuscript. This work was partially funded by the Danish National Research Foundation (DNRF) through establishment of the Discovery Center and the Villum Fonden through the Deep Space project. Hao Liu is supported by the National Natural Science Foundation for Young Scientists of China (Grant No. 11203024) and the Youth Innovation Promotion Association, CAS.

\ifx\usePRL\defined
\end{acknowledgments}
\fi

\end{document}